\documentclass[a4paper,11pt]{article}
\usepackage{amsmath,graphicx,color,mathrsfs,subfigure,multicol}

\def\circa#1{\,\raise.3ex\hbox{$#1$\kern-.75em\lower1ex\hbox{$\sim$}}\,}
\makeatletter

\def\art{\@ifnextchar[{\eart}{\oart}}
\def\eart[#1]#2#3#4#5#6{{\rm #2}, {\em #3  #4} {\rm (#6) #5} ({\em #1})}
\def\hepart[#1]#2{{\rm #2, \em#1}}
\newcommand{\oart}[5]{{\rm #1}, {\em #2  #3} {\rm (#5) #4}}

\newcounter{alphaequation}[equation]
\def\thealphaequation{\theequation\hbox to
0.6em{\hfil\alph{alphaequation}\hfil}}
\def\eqnsystem#1{
\def\@eqnnum{{\rm (\thealphaequation)}}
\def\@@eqncr{\let\@tempa\relax \ifcase\@eqcnt \def\@tempa{& & &} \or
  \def\@tempa{& &}\or \def\@tempa{&}\fi\@tempa
  \if@eqnsw\@eqnnum\refstepcounter{alphaequation}\fi
\global\@eqnswtrue\global\@eqcnt=0\cr}
\refstepcounter{equation} \let\@currentlabel\theequation \def\@tempb{#1}
\ifx\@tempb\empty\else\label{#1}\fi
\refstepcounter{alphaequation}
\let\@currentlabel\thealphaequation
\global\@eqnswtrue\global\@eqcnt=0 \tabskip\@centering\let\\=\@eqncr
$$\halign to \displaywidth\bgroup \@eqnsel\hskip\@centering
$\displaystyle\tabskip\z@{##}$&\global\@eqcnt\@ne
\hskip2\arraycolsep\hfil${##}$\hfil& \global\@eqcnt\tw@\hskip2\arraycolsep
$\displaystyle\tabskip\z@{##}$\hfil
\tabskip\@centering&\llap{##}\tabskip\z@\cr}
\def\endeqnsystem{\@@eqncr\egroup$$\global\@ignoretrue} \makeatother

\begin{document}
\begin{flushright}
{hep-th/0603026} \\
UMN-TH-2433/06\\
\end{flushright}
\vspace{0.5cm}

\begin{center}
{\LARGE \bf
Standard $4$d gravity on a brane in six dimensional\\ flux compactifications}\\[1cm]

{
{\large\bf Marco Peloso$^{a\,}$\footnote{E-mail:  peloso@physics.umn.edu}
\,,\,  Lorenzo Sorbo$^{b\,}$\footnote{E-mail:  sorbo@physics.umass.edu}
\,,\, Gianmassimo Tasinato$^{c\,}$\footnote{E-mail:  tasinato@thphys.ox.ac.uk}
}
}
\\[7mm]
{\it $^a$ School of Physics and Astronomy,
University of Minnesota, Minneapolis, MN 55455, USA}\\[3mm]
{\it $^b$Department of Physics, University of Massachusetts, Amherst,
MA 01003,
USA
}\\[3mm]
{\it $^c$ The Rudolf Peierls Centre for Theoretical  Physics,
 Oxford University, Oxford OX1 3NP, UK
}
\\[1cm]
\vspace{-0.3cm}

\vspace{1cm}

{\large\bf Abstract}

\end{center}
\begin{quote}

{
We consider a six dimensional space-time, in which two of the dimensions are compactified by a flux. Matter can be localized on a codimension one brane coupled to the bulk gauge field and wrapped around an axis of symmetry of the internal space. By studying the linear perturbations around this background, we show that the gravitational interaction between sources on the brane is described by Einstein $4$d gravity at large distances. Our model provides a consistent setup for the study of gravity in the rugby (or football) compactification, without having to deal with the complications of a delta--like, codimension two brane. To our knowledge, this is the first complete study of gravity in a realistic brane model with two extra dimensions, in which the mechanism of stabilization of the extra space is fully taken into account.
}

\end{quote}


\def\der{\partial}
\def\fk{f_\kappa}
\def\t{t_b}
\def\r{r_b}
\def\c{\chi_b}
\def\G{\Gamma}
\def\Bt{{B_\theta}}
\def\Bp{{B_\phi}}
\def\bt{{\bar \theta}}
\def\hp{{\hat \phi}}
\def\tg{{\rm tg }\,}
\def\hmn{h_{\mu \nu}}
\def\cmn{C_{\mu \nu}}
\def\e{{\rm e}}
\def\be{\begin{equation}}
\def\ee{\end{equation}}
\def\bea{\begin{eqnarray}}
\def\eea{\end{eqnarray}}
\def\ni{\noindent}

\section{Introduction}

Models with two extra dimensions have received 
large attention, both in the past and today.
Among them, there are the first examples
of supersymmetric 
flux compactifications to four dimensional
Minkowski space.  They are characterized by a magnetic monopole 
that spontaneously
compactifies two of the dimensions on a sphere \cite{ss}, stabilizing
the internal space. Models in this class
have also been  considered to build
brane-world scenarios \cite{add}: two
is the minimum number of flat extra dimensions to have a fundamental  scale
of order  TeV, and a compactification radius of (sub)millimeter range.
   Field theory models with two extra dimensions were studied, starting
with \cite{sundrum}, due to their interesting properties
that differentiate them from the codimension one case \cite{fieldtheory6}.
 Also,  the behavior of gravity
in brane models with two, warped or unwarped, extra dimensions have been
analyzed 
in specific cases \cite{gherghetta}, revealing interesting
features in comparison with codimension one.

More  recently, the models \cite{ss} in which two extra dimensions
are compactified by fluxes, have been reconsidered in a brane-world
scenario, in order to address the cosmological
constant problem \cite{others}. In this new 
 approach, the original spherical
compactification is modified by cutting a wedge to the sphere, and identifying the two
edges giving a geometry with the shape of a rugby-ball (or a football one, according
to the geographical location of the reader). This  introduces
a deficit angle on the sphere, with two conical singularities at the poles of the
compactification manifold.
The two conical singularities can be interpreted as codimension two branes,
in which matter is supposed to be localized.
Even  without discussing the relevance
of this approach for the  cosmological constant problem (see~\cite{others,gapo} for a critical discussion)
this scenario is interesting by its own.
It provides a natural setting to study brane-world models in which  
  extra-dimensions are compactified, and possibly 
  stabilized, by fluxes.

\smallskip

In the present paper we study the large distance behavior of gravity in a
brane-world  scenario with flux  compactification. To do this, we study how linear perturbations
of the brane energy-momentum tensor affect the bulk geometry, and the
background gauge field.
The first problem one meets, on facing
these questions, is that 
  energy momentum tensor
different
from pure tension {\it cannot} 
 be  accommodated on strict  codimension two
singularities, in the context of Einstein gravity  \cite{cline,gregory}.
 A way out to the problem has been proposed by~\cite{gregory,corradini}, through the addition of Gauss-Bonnet terms in the bulk; a solution where gravity is less dramatically modified could be to promote the brane to a thick defect: this is the approach taken in this paper. An apparent problem with this approach
is the fact that it is model dependent, since
 there are various inequivalent
ways to regularize a conical singularity \cite{geroch}.  
On the other
hand, we believe, our construction is the simplest and the most natural one, and has
 interesting features that
 render its analysis worth of attention.
 
 \smallskip

In our model, the strict codimension two defect
is substituted by a codimension one brane, with one spatial
dimension compactified on a circle. The bulk space-time corresponds
to a regularized version of the rugby-ball compactification, in which
the tip of the conical singularities are substituted by regular spherical
caps that smoothly end the space. The spherical caps are joined
to the rugby ball region at the position of the branes. The latter
must satisfy proper junction conditions to compensate possible jumps
on the derivative of the metric and the gauge potentials.
Precisely for this reason,  the brane is   coupled to
the bulk magnetic field, and, given our Ansatz for the metric,
 this coupling fixes its position
in the bulk at the background level.
 Consequently, the very same bulk magnetic
 field that compactifies
and stabilizes the extra dimensions, also stabilizes the
brane position. 

Due to the presence of the bulk gauge  field, our model allows to study,
in a simple setting, some of the features of brane models in string theory, in which
moduli are stabilized  by fluxes. Using a thick defect, we naturally
avoid the pathologies of the strict
codimension two case, and we are allowed  to place our preferred
energy momentum tensor on the brane. 
An additional
motivation for our scenario comes from field theory models, in which
thick codimension two branes in six dimensions have been considered
\cite{dudas}. In that context, the thickness of the brane
is used as cutoff for certain regularization group flows. In these models,
a thick brane is not just a regularized version of a (more fundamental) 
strict codimension two object: on the contrary, the brane thickness
plays a crucial role in the discussion~\footnote{
Also, configurations containing thick codimension two branes
have been found  looking for  vacua of six dimensional
gauged supergravity \cite{thickvacua}.}.

\smallskip

After presenting our background,
 we proceed with the study of 
 linear perturbations on it.  These are necessary to study the 
backreaction of brane matter on the geometry and the bulk fields, 
and to analyze the behavior of gravity on the brane.  
  Analysis of the behavior of gravity and cosmology in
codimension two brane-worlds  have already been considered
in the literature \cite{precedenti1,precedenti2} (see for
example \cite{wilt} for a study of non-linear
gravitational waves propagating on a codimension one brane
in a regular six dimensional background).  Sometimes, previous 
studies found 
  that the energy momentum tensor on the brane must satisfy precise
  conditions in order to recover standard cosmology at late times.
  In the present paper
we  consider  the most general Ansatz for
the perturbations, and  we  consistently 
include  the necessary couplings between
the bulk gauge field and the brane. 
At the linearized level, the system of equations that governs
the massless  perturbations can be solved exactly. All the scalar quantities depend on two functions, that control the bulk geometry, the gauge field, and the position of the brane in the bulk. 
By studying the junction conditions for the metric and the gauge field, 
we can provide a clear
geometrical interpretation  of the effect of matter on the brane.
The brane is bent with respect to its initial position
(like in the codimension one case \cite{gartan}) and the
geometry of the internal space gets deformed.

\smallskip

The natural  application of our analysis is the study of the behavior
 of induced gravity on the brane. 
Various massless scalar modes 
contribute to gravitational equations, potentially inducing large-distance
corrections to Einstein gravity. 
We show that all the equations
reorganize in such a way that Einstein
  gravity is reproduced at 
large distances on the brane.
  In particular, we recover the standard Einstein
equations in four dimensions, plus
 corrections due the field that controls
 the compactification volume. However,  using the junction conditions,
we are able to show that 
these corrections are 
sub-dominant with respect to the standard contributions, and become important 
only at scales comparable to the size of the compactification manifold.
Having done so, we are finally able to compute the limit in which the brane shrinks to a codimension-two defect. This computation is particularly important, since it can shed light on the behavior of gravity when coupled to higher codimension branes. We discuss this issue in the conclusions.

\smallskip

The paper is organized as follows. In Section 2 we present the background
configuration for our model. The study of linear perturbations
is started in Section 3. In Section 4 we discuss the solutions
for the equations ruling 
the  perturbations, and we impose the necessary junction conditions
at the brane position. Section 5 is devoted to study the behavior
of gravity as seen by a brane observer. Section 6 contains 
 our conclusions. 
A considerable number of Appendixes is added at the end of the paper,
 in order
to  provide the necessary tools to 
reproduce the results discussed in the main text.

\section{The model and the background solution}
\label{backsec}

The action of the model is
\begin{eqnarray}
S &=& S_o + S_i + S_{\rm str} \nonumber\\
S_{\rm o,i} &=& \int d^{6}x \sqrt{-g_{6}} \left[ M^{4}\, {\mathcal R}-\Lambda_{\rm o, i}
-\frac{1}{4} F_{A B} F^{A B} \right] \label{actionbulk} \\
S_{\rm str} &=& - \int d^5 x \sqrt{-\gamma} \left[ \lambda_s
+\frac{v^2}{2} \left(\partial_M\sigma-{\mathrm e}\,A_{M}\right)\,\left(\partial^M\sigma-{\mathrm e}\, A^{M}\right)  \right]
\label{actionbrane}
\end{eqnarray}
Let us discuss separately the two sectors, bulk and brane.

\bigskip

\noindent
{\it The bulk}

\bigskip

\noindent
Let us start by describing the background bulk solution. We consider Minkowski spacetime times a two dimensional compact space. The latter resembles the so called ``rugby ball'' compactification, characterized by azimuthal symmetry, a deficit angle, a $Z_2$ symmetry across the equator, and two conical singularities at the poles. It is convenient to use polar coordinates for the internal space; however, we use a slightly unconventional origin for the polar angle $\theta$, placing the equator at $\theta = 0 \,$, and the two poles at $\theta = \pm \pi / 2 \,$ (so that the $Z_2$ symmetry is simply $\theta \leftrightarrow - \theta \,$). We still impose the $Z_2$ symmetry, so that  it is enough to describe the ``upper'' region, $\theta \geq 0 \,$. At odds with the standard construction, where a codimension two brane is placed at the conical singularity, we truncate the rugby ball at the fixed angle angle $\theta=\bar{\theta}$, and we terminate the space with a regular spherical cap with no deficit angle. The two regions are joint through a codimension one cylinder $S^1\times$(4d Minkowski). The defect can be also viewed as a string in the internal space (once the noncompact directions are suppressed), and, therefore, the term string will be also used in the following.

\begin{figure}[h]
\centerline{
\includegraphics[width=0.25\textwidth]{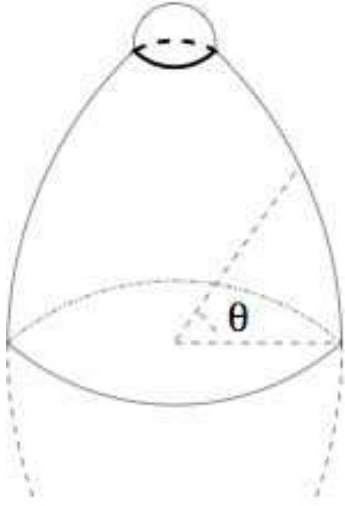}
}
\caption{The background configuration.}
\label{fig:fig1}
\end{figure}

We denote the region from the equator to the string as the ``outside bulk'', and the region from the string to the pole as the ``inside bulk''; the overall geometry is sketched in fig.~\ref{fig:fig1} (we stress that, while a $Z_2$ symmetry is imposed at the equator, there is no the $Z_2$ symmetry across the string).

\smallskip
\ni
The outside bulk, located at $0< \theta < \bar{\theta}$, is characterized by
\bea
d s_{6}^{2} &=&   \eta_{\mu \nu} d x^{\mu} d x^{\nu} +R_{o}^{2}\, d
\theta^{2}+R_{o}^{2} \,\beta_{o}^{2} \cos^{2}{\theta} \, d
\hat{\phi}^{2} \\
F_{\theta  \hat{\phi}} &=&   M^{2} \beta_{o} R_{o}\, \cos{\theta}\,\,.
\eea
where $R_o$ denotes the compactification radius, and $1-\beta_o$ is the deficit angle.

For the interior geometry, $\bar{\theta} < \theta <\frac\pi2$, we have instead
\bea
d s_{6}^{2} &=&   \eta_{\mu \nu} d x^{\mu} d x^{\nu} +R_{i}^{2}\, d
\theta^{2}+R_{i}^{2} \,\beta_{i}^{2} \cos^{2}{\theta}
 \, d \hat{\phi}^{2} \\
F_{\theta \hat{\phi}} &=&  M^{2} \beta_{i} R_{i}\,  \cos{\theta}\,\,.
\eea
(where $R_i$ and $\beta_i$ are the analogous of $R_o$ and $\beta_o \,$, respectively). 

This compactification is obtained by two different six dimensional cosmological constants $\Lambda_{o,i}$,
\begin{equation}
\sqrt{2 \lambda_{i}} \,=\, \frac{M^{2}}{R_{i} }\qquad\qquad
\sqrt{2 \lambda_{o}} \,=\,
\frac{M^{2}}{R_{o} }\,.
\end{equation}
and by the two different values of the ``magnetic'' field $F_{\theta {\hat \phi}}$ given above.

As mentioned, we choose the parameters to have a regular inside bulk with no deficit angle, $\beta_{i}=1$. The continuity of $g_{\hat{\phi}\hat{\phi}}$ at the brane position then imposes
 \be
 R_{i}=\beta_{o} R_{o}
 \ee

A few redefinitions allow to write the bulk solution in a more compact form. We redefine $R\equiv R_{o}$ and $\beta= \beta_{o}$, and introduce the dimensionful coordinates $l ,\, \phi \,$, satisfying
\begin{eqnarray}
d l &=&  R \, \left[ \Theta \left( \bt -  \theta  \right) + \beta \, \Theta
\left( \theta -\bt \right) \right] d \theta \,\,,
\nonumber\\
d \phi &=& R \, \beta \, d \hat{\phi} \,\,.\label{change}
\end{eqnarray}
where $\Theta$ is the Heaviside step function. In terms of these coordinates the bulk solution rewrites
\begin{eqnarray}\label{nfsol}
d s^2 &=& d l^2+
\cos^2\left[ \theta \left( l \right)\right]\,d\phi^2
+\eta_{\mu\nu}\,
dx^\mu\,dx^\nu \\
F_{l \phi} &=& M^{2} \, \theta' \left( l \right)\,\cos
\left[ \theta \left( l \right)\right]\label{intrgf}
\end{eqnarray}
both outside and inside (here and in the following prime denotes differentiation with respect to $l \,$). The function $\theta \left( l \right)$ is obtained by inverting~(\ref{change}) 
\begin{equation}
\theta \left( l \right) = \bt + \frac{1}{R}
\left(  l - R \, \bt \right) \left[  \Theta \left( R\,\bt-  l \right) + \frac{1}{\beta} \, \Theta \left(l -R\,\bt \right)
 \right] \,\,.
\end{equation}
In the following we will need some properties of this function:
\begin{eqnarray}
\theta'(l) &=& \frac{1}{R}\left[ \Theta \left( R \, \bt -  l
 \right) +
\frac{1}{\beta} \, \Theta \left(  l - R \, \bt \right) \right]\,,\\
\theta''(l) &=& - \frac{1}{R} \left(1-\frac{1}{\beta}\right)\,\delta(l
-R \bt )\,.
\end{eqnarray}
Moreover, we see that the two new coordinates extend for
\begin{equation}\label{extcoord}
0 < l < R \left[ \bt + \beta \left( \frac{\pi}{2} - \bt \right) \right] \;\;\;,\;\;\;
0 < \phi < 2 \, \pi \, R \, \beta \,\,,
\end{equation}
with the string at $l = R \, \bt \,$. 

We conclude by noting that if some field, with charge $\e$ under the $U (1)$ symmetry, is present, it is well known that the deficit angle must satisfy a quantization condition
\begin{equation}
\beta = \frac{N}{ 2\,\e\, M^2 R} \;\;\;,\;\;\; N = 0 ,\, 1 ,\, 2 ,\, \dots \,\,.
\label{quantization}
\end{equation}
as we recall in Appendix~\ref{appA}. At least one charged field must be present on the brane (see the next  paragraph) and so this condition holds for our system.

\bigskip

\noindent
{\it The brane}

\bigskip

\noindent
The discontinuities in the geometry and in the gauge field must be compensated by energy momentum tensor localized  on a codimension one brane, at the position $ l = R \, \bt \,$. To compensate the jump in the Maxwell equations (discussed below), the brane must couple also
to the gauge field. This motivates the choice of the brane
action (\ref{actionbrane}) \cite{otto}. In particular, we interpret $v/\sqrt{2}$ as the absolute value of the 
vacuum expectation value of 
a brane Higgs field of charge ${\mathrm e}$, that we have integrated out. 
$\sigma$ is the phase of such a field, and acts as a Goldstone mode. 
The  induced metric on the four brane is given by
 \be
 d s_{5}^{2}=
\eta_{\mu \nu}\,d x^{\mu} d x^{\nu}+ \cos^2{\bar{\theta}} \,d \phi^{2}
 \ee
The energy momentum tensor of the four  brane is
\begin{eqnarray}\label{brbackenten1}
S_{\mu \nu } &=& -\left[\lambda_{s} +\frac{v^2}{2}\,\left(
\partial_\phi \sigma-{\mathrm e}\,A_{\phi}\right)\,
\left(\partial^\phi\sigma-{\mathrm e}\,A^{\phi}\right) \right] \,\eta_{\mu \nu} \\
S_{\phi \phi} &=&  -\left[\lambda_{s}- \frac{v^2}{2}\,\left(
\partial_\phi \sigma-{\mathrm e}\,A_{\phi}\right)\,
\left(\partial^\phi\sigma-{\mathrm e}\,A^{\phi}\right) \right] \,\gamma_{\phi\phi}\label{brbackenten2}
\end{eqnarray}
while $S_{\mu \phi} = 0 \,$.

We must solve the Israel junction conditions
\begin{equation}
\left[ \hat{K}_{A B}\right]_{J}=-\frac{S_{A B}}{M^4}\,.
\label{israel}
\end{equation}
where
\begin{equation}
\hat{K}_{M N}= K_{M N}-\gamma_{M N}K\,\,\,\,,\,\,\,\,
K_{M N}=\nabla_{M} n_{N}
\end{equation}
and where we have defined  $[f]_{J}\equiv f_{out}-f_{in}$.

For the above background, one finds
\begin{equation}
\left[ \hat{K}_{\mu \nu} \right]_{J}=
\frac{1-\beta}{R\beta} \tan{(\bar\theta)} \,\eta_{\mu\nu}\,\,,\,\,\,\,\,
\hat{K}_{\phi\phi}=0\,\,.
\end{equation}
So $S_{\phi \phi}$ must be zero, which fixes $\lambda_s$ in terms of $v$, $\sigma$ and $\e$:
\begin{equation} 
\lambda_{s}= \frac{v^2}{2}\,\left(
\partial_\phi \sigma-{\mathrm e}\,A_{\phi}\right)\,
\left(\partial^\phi\sigma-{\mathrm e}\,A^{\phi}\right)
\label{reltosaj}
\,.\end{equation}
The equation of motion for $\sigma$, $\partial^M \left(\partial_M\sigma-{\mathrm e}\,A_M\right)=0$, implies $\partial^\phi\,\partial_\phi\sigma=0$. This latter equation has to be solved while taking into account that $\sigma$ is a phase and that the allowed values for $\beta\,R$ are related to ${\mathrm e}$ by eq.~(\ref{quantization}). This gives (up to an irrelevant additive constant)
\begin{equation}\label{solsigma}
\sigma\left(\phi\right) = \frac{n}{R \beta} \phi = \frac{2\,n}{N}\,{\mathrm e}\,M^2\,\phi
\end{equation}
where $n$ is an integer.
From the Maxwell equation, one  obtains the following junction condition
for the discontinuity  of the magnetic field
\begin{eqnarray}\label{jumpmaxw}
\left[ \left(\sqrt{g_{ll}} F^{l \phi} \right)\right]_{J}
 ={\mathrm e}\, v^{2}\, 
\left(\partial^\phi\sigma-{\mathrm e}\,A^\phi\right)_{|_{brane}}
\end{eqnarray}

As shown in appendix~\ref{appA}, $A_\phi{}_{|_{brane}}=M^2\,\left(\sin\bt-1\right)$. Then, eq.~(\ref{jumpmaxw}) and the remaining ($\mu\nu$) component of~(\ref{israel}) give
\begin{eqnarray}
&& n \,=\, - N / 2 \\
&& \frac{1}{R} \left(1-\frac{1}{\beta} \right) \cos \bt \,=\,  -q^{2} \, \sin \bt \label{fixbr1}\,\,,\quad\qquad q\,\equiv \,{\mathrm e}\,v\,\,.
\end{eqnarray}
Notice that the first relation forces us to consider only systems where $N$ is even.

To summarize, after the mode $\sigma$ is set to its solution, the brane action is characterized by the three parameters $q^2$, $\lambda_s$, and $\bt$. The brane position $\bt$ is related to the (rescaled) charge $q^2$ by eq.~(\ref{fixbr1}). Eq.~(\ref{reltosaj}) then gives the tension $\lambda_s \,$. This last relation can be rewritten as
\begin{equation}
 q^{2}\,=\,-\left[\theta'\right]_{J} \,\frac{1}{\tan{\bar
\theta}}\,=\,
\frac{2 \lambda_{s}}{M^4\,\tan^2{\bar{\theta}}}\label{defq2}
\end{equation}

It is instructive to compare these result to the expression for the deficit angle which is obtained in the case of a codimension $2$ object localized at the pole~\cite{others}
\begin{equation}
1 - \beta = \frac{T}{2 \pi M^4} \;\;,\;\; {\rm codimension } \; 2
\label{deficitcod2}
\end{equation}
\smallskip
\noindent

In our case, the total tension on the string is obtained by integrating
over the azimuthal direction the $(\mu\nu)$ component of the
energy momentum tensor on the string
\begin{equation}
T = 2 \pi R \beta\, \cos{\bt}\,\left[
\lambda_{s}+\frac{v^2}{2}\left(\partial_\phi\sigma-{\mathrm e}\, A_{\phi}\right)\,\left(\partial^\phi\sigma-{\mathrm e}\, A^{\phi}\right) 
\right]
\label{tottens}
\end{equation}
Inserting the
expressions we have
determined, the deficit angle in our case reads,
\begin{equation}
1 - \beta = \frac{T}{2M^4 \pi \sin{\bt}}
\end{equation}
If we now shrink the string to the north pole (i.e. $\bt\rightarrow \pi/2$), we recover the codimension $2$ result~(\ref{deficitcod2}).

\section{Perturbations}

The defect has the topology of a cylinder, once the non-compact directions are also taken into account.
  Fields localized on the defect have zero modes which appear as ``rings'' on this 
cylinder, with the wave function homogeneously distributed along the compact $\phi$ direction. We
will be
 interested in computing gravity at large distance between two zero modes localized on the defect. The bulk gravitons which give a dominant contribution to the interaction will also be symmetric in $\phi \,$. For this reason, we can restrict our study to azimuthally symmetric perturbations of the background solution~(\ref{nfsol}).
   Since we are interested in gravity at distances much larger than $R \,$, in the following we further restrict
 our analysis  to the massless bulk modes.

The most general geometry with this property is described by
\begin{eqnarray}
ds^2&=&\left(1+2\,\Phi\right)\,d l^2+2\,A\,d l\,d\phi+
\left(1+2\,C\right)\,\cos^2 \theta \,
d\phi^2\nonumber\\
&+& 2 \left( T_\mu + \partial_\mu T \right) d l d x^\mu + 2 \left( V_\mu + \partial_\mu V \right) d \phi \, d x^\mu \\
&+&\left\{\eta_{\mu\nu}\,\left(1+2\,\Psi\right)+2\,E,_{\mu\nu}+ E_{(\mu,\nu)} +
\hmn\right\}\,dx^\mu\,dx^\nu \nonumber
\label{perturbedds}
\end{eqnarray}
where none of the fields depend on $\phi \,$, and where $E_{(\mu \nu)} \equiv \partial_\nu E_\mu + \partial_\mu E_\nu \,$.
We have split the perturbations into scalar, vector and tensor modes under 
the Lorentz group
in the non-compact coordinates. The vector modes $T_\mu ,\, V_\mu ,\, E_\mu$ are transverse, $\partial^\mu T_\mu = 0 ,\, \dots$, while the tensor mode $h_{\mu \nu}$ is transverse and traceless, $\partial^\mu h_{\mu \nu} = 0 \;,\; h \equiv h^\mu_\mu = 0 \,$. The remaining modes are scalar. This choice is motivated by the fact that modes belonging to different representations are not coupled to each other at the linearized level (this considerably simplifies the study of the Einstein equations in the bulk).

The modes~(\ref{perturbedds}) transform under infinitesimal change of coordinates. As we discuss in details in appendix~\ref{appB}, we can fix (part of) the gauge freedom, and reduce the number of modes by imposing $E_\mu = T = V = 0 \,$. This leaves the residual freedom
\begin{equation}
x^A \rightarrow x^A + \xi^A \left( x^\mu ,\, l \right) \;\;\;{\rm with}\;\; \xi^l = - \xi' ,\; \xi^\phi = 0 ,\, \xi^\mu = \partial^\mu \xi
\label{residual}
\end{equation}
We further restrict the gauge freedom by imposing that 
\begin{equation}
\theta_{\rm string} = \bt \;\;\;,\;\;\; \Phi \left( \bt \right) = 0 
\label{gn}
\end{equation}
that is we require that the string lies at the unperturbed background position, and that $g_{ll} = 1 \,$ there. This choice includes the Gaussian normal coordinates, but we impose this only at the brane, and not necessarily for a finite distance (along $\theta$) from the brane. For shortness, we still refer to~(\ref{gn}) as choosing gaussian normal coordinates, although it is actually more general.
 The choice~(\ref{gn}) forces $\xi^l \left( \bt \right) = \xi^{l}{}' \left( \bt \right) = 0 \,$, leaving however the residual freedom of any function $\xi \left( x^\mu \right) \,$ (this freedom is used below when we compute the gravitational interaction). We rather write the bulk equations in a manifestly gauge invariant form by identifying gauge invariant modes under~(\ref{residual}), as was originally done for the cosmological $4$d case in~\cite{bardeen}. As we show in appendix~\ref{appB}, the remaining vector and tensor modes, as well as the scalars $A$ and $\Psi$ are already invariant under~(\ref{residual}), while the other scalar modes combine into the gauge invariant combinations
\begin{eqnarray}
{\hat \Phi} &\equiv&  \Phi + E'' \nonumber \\
{\hat C} &\equiv&  C - \theta' \tan \theta \, E'
\label{invariant1}
\end{eqnarray}

In an arbitrary gauge, the string would be at position
\begin{equation}
\theta_{\rm string} = \bt + \zeta \left( x^\mu \right)
\end{equation}
The mode $\zeta$ corresponds to a ($\phi$ symmetric) deformation of the defect, and it can be thought of as the analogous of the ``brane bending'' mode of~\cite{gartan}. We show below that, also in the present context, the mode is crucial to obtain standard $4$d gravity at large distance. The brane position changes under the residual coordinate transformation~(\ref{residual}). However, it can be combined to give the gauge invariant displacement
\begin{equation}
{\hat \zeta} = \zeta - E'
\end{equation}
Therefore,
since we choose
$\zeta=0$,
 the deformation of the string is encoded in $- E' \left( \bt \right) \,$.

The field $\sigma$ is localized on the brane; we denote its perturbation by $\delta\sigma$, that can be shown to be invariant under the residual coordinate transformation~(\ref{residual}).

Finally, we must also consider the perturbations of the gauge field,
\begin{equation}
\delta A_\phi \equiv a_\phi ,\; \delta A_l \equiv a_l ,\; \delta A_\mu \equiv \partial_\mu a + {\hat a}_\mu
\label{pertgauge}
\end{equation}
where $\delta A_\mu$ has also been decomposed into a scalar and a transverse vector part. We denote the corresponding perturbations of the field strength as $\delta F_{A B} \equiv f_{AB} \,$. The $f_{l \mu}$ component receive contributions from both modes in the vector and scalar sectors,
\begin{equation}
f_{\mu l} = \der_\mu \left( a_l - a' \right) - {\hat a}_\mu' \equiv f_{\mu l}^S + f_{\mu l}^V \,\,,
\end{equation}
The two terms are not separately invariant under $U \left( 1 \right)$ transformations. As we discuss in
appendix~\ref{appB}, we use (part of) the gauge freedom to set $f_{\mu l}^S = 0 \,$ in the bulk.

Only the component $a_\phi$ transforms under the residual coordinate transformation~(\ref{residual}). Also in this case, we can construct the gauge invariant combination
\begin{equation}
{\hat a}_\phi = a_\phi + \theta' M^2 \cos \theta \, E'
\label{invariant2}
\end{equation}

\section{Zero mode solutions and localization of sources on the string}

We restrict to zero modes since they are the only ones responsible for gravity at large distances. The physical modes obey several equations, both in the bulk and at the string. We first solve the Einstein and
Maxwell equations in the bulk (the details of the computation are shown in appendix~\ref{appC}). These gives the bulk solutions in terms of several integration constants (the computation is performed on the outside and the inside separately). Some of these modes are projected out by requiring appropriate $Z_2$ parity at the equator, and regularity at the pole.

The integration constants are then related by a number of boundary conditions at the string. We first impose the first Israel conditions (appendix~\ref{appD}), which are the continuity of the induced metric at the brane
\begin{eqnarray}
\gamma_{\phi \phi} &=& \cos^2 \bt \left( 1 + 2 C \right) \nonumber\\
\gamma_{\mu \nu} &=& \left( 1 + 2 \Psi \right) \eta_{\mu \nu} + E_{,\mu\nu} + h_{\mu \nu}
\label{induced}
\end{eqnarray}
(that is, the above components should be equal for $\theta \to \bar{\theta}^{\pm}$).
 The function $E$ at the string does not play any role, and indeed it can be gauged to zero through the residual coordinate transformation $\xi \left( x^\mu \right) \,$. We also impose continuity of the $\mu$ and $\phi$ components of the gauge field $A^M$ (otherwise the string action~(\ref{actionbrane}) would not be properly defined). In Appendix~\ref{appE} we then solve the Maxwell equations at the brane.
 
Matter on the string (assumed to be uncharged under the bulk gauge group) does not enter in any of the equations listed so far. Therefore, it is worth summarizing the general results up to this point, before introducing matter fields.

\smallskip

The tensor mode is forced to be $\theta-$independent, and equal on the outside and the inside bulk,
\begin{equation}
h_{\mu \nu} = C_{\mu \nu} \left( x \right) \;\;\;,\;\;\; {\rm everywhere}
\label{hsol}
\end{equation}

\smallskip

The vector mode $T_\mu$ is forced to vanish; the other modes $a_\mu$ and $V_\mu \,$ are not coupled to matter (they do not contribute to the junction conditions) at the linearized level, and we therefore do not need to consider them further. 

\smallskip

The scalar mode $A$ is forced to vanish. The brane scalar mode $\delta\sigma$ is also forced to depend only on $x^\mu$ once we impose the periodicity on  $\sigma+\delta\sigma$ (that is, $\sigma+\delta\sigma \rightarrow \sigma+\delta\sigma + 2 \pi n$ as $\phi \rightarrow \phi + 2 \pi R \beta \,$) and we take into account that $\delta \sigma$ is infinitesimal (so that it cannot contribute to this finite discontinuity). The combination $a_l - a'$ is set to zero by a $U \left( 1 \right)$ gauge choice (see the discussion around eq.~(\ref{u1fs})). The modes enter separately only in the brane equations (where the $U \left( 1 \right)$ gauge symmetry is broken); the boundary conditions give $a_l = 0$, $\delta \sigma= e \, a$ at the brane and we can set $a\left(\bt\right)=0$ by using a residual $U\left(1\right)$ gauge freedom (again, cf. the discussion after eq.~(\ref{u1fs})). Therefore, also these modes can be discharged. The remaining modes must be included. Due to the residual freedom~(\ref{residual}), the bulk equations can be written in terms of only the gauge invariant combinations ${\hat \Phi} ,\, {\hat C} ,\, \Psi ,\, {\hat a}_\phi \,$. The brane equations can be also written in terms of these combinations, plus the gauge invariant brane mode~\footnote{More precisely, we do not have a definite parity across the string, and the brane displacement in the outside and inside bulk is controlled by the two different gauge invariant modes, ${\hat \zeta}_{\rm out} ,\, {\hat \zeta}_{\rm in} \,$. A $Z_2$ symmetry, as for instance in the RS case, would force these mode to be one the opposite of the other, but this is not true in the present case.} $\, {\hat \zeta} = - E' \,$. As we show in appendices~\ref{appB} through~\ref{appE}, all these modes can be written in terms of two $x^\mu-$dependent integration constants, corresponding scalar fields in $4$d. Schematically,
\begin{equation}
C_\Psi^{(i)} \left( x \right) ,\, D_\Psi^{(i)} \left( x \right) \;\;\rightarrow\;\;
{\hat \Phi} ,\, {\hat C} ,\, \Psi ,\, {\hat a}_\phi ,\, {\hat \zeta}
\end{equation}

\noindent
The suffixes for the integration constants are justified by the fact that the function $\Psi$ on the inside part of the bulk depends on both the modes in the following simple way
\begin{equation}
\Psi \left( \theta ,\, x^\mu \right) = C_\Psi^{(i)} \left( x \right) + D_\Psi^{(i)} \left( x \right) \, \sin \theta \;\;\;\;,\;\;\;\; {\rm inside}
\label{name}
\end{equation}
(outside, instead, $\Psi$ is forced to be constant, and, by continuity, equal to $C_\Psi^{(i)} + D_\Psi^{(i)} \sin \bt \,$). So, considering all but the (second) Israel conditions, we end up with a tensor ($C_{\mu \nu})$ and two scalar ($C_\Psi^{(i)} \,, D_\Psi^{(i)}$) zero modes.

\bigskip

The interaction of these modes with matter on the brane is described by the (second Israel) junction conditions (see appendix~\ref{appF})
\begin{eqnarray}
\left[ - {\hat \zeta}_{,\mu\nu} + \frac{1}{2} h_{\mu \nu}' \right]_J &=&
\frac{1}{M^4} \left( T_{\mu \nu} - \frac{T}{3} \eta_{\mu \nu} \right) \label{j-munu} \\
\left[ \partial^2 {\hat \zeta} - \frac{4 \Psi'}{\cos^2 \bt} \right]_J &=&
\frac{T_{\phi}^{\phi}}{M^4} \label{j-phiphi}
\end{eqnarray}
where $T_{\mu \nu}$ and $T_{\phi \phi}$ are the non vanishing components of the energy--momentum tensor of matter fields on the string. These equations allow to determine the three zero modes of the system. 
 
Let us observe that the trace of~(\ref{j-munu}) gives the simple relation
\begin{equation}
\left[ \partial^2 {\hat \zeta} \right]_J = \frac{T}{3 M^4} \;\;\;,\;\;\; T \equiv T^\mu_\mu
\label{trace}
\end{equation}
between the deformation mode of the string and the trace of the matter energy--momentum tensor
along the non-compact directions. From this, eq. (\ref{j-phiphi}) rewrites
\begin{equation}
\left[  \frac{4 \Psi'}{\cos^2 \bt} \right]_J =\frac{T}{3 M^4}-
\frac{T_{\phi}^{\phi}}{M^4}
\label{psijtt}
\end{equation}

\noindent
Being $\left[ \Psi'\right]_J \propto D_{\Psi}^{(i)}(x)$, we discover that, if the energy--momentum tensor on the string satisfies
\begin{equation}
\frac{T}{3} = T_{\phi}^{\phi}
\label{finetune}
\end{equation}
the massless mode $D_{\Psi}^{(i)}$ is removed from the spectrum.  As we discuss below, this mode controls the variation of the volume of the internal space (which is instead independent of $C_\Psi^{(i)} \,$), and leads to a deviation (however negligible, at large distances) from standard Einstein $4d$ gravity. Note that the condition~(\ref{finetune}) is analogous to the condition  $T^5_5=T/2$ given in~\cite{kanti} for stability of codimension--one brane models.
 
\section{Gravity at large distance}

We are now in the position to compute gravity at large distance ($\gg R$) between two sources localized on the string. Most of the computation 
extends the analogous one of~\cite{gartan} for gravity in the Randall--Sundrum model~\cite{rs2}, taking into account now   extra--complications 
 due to the presence of the $\phi$ coordinate and the additional
bulk fields. The first step is to notice that the bulk equation for the tensor modes, eq.~(\ref{eint}), and the junction condition~(\ref{j-munu}) can be combined into the single equation
\begin{equation}
\partial_l \left( \cos \theta \, \partial_l \, h_{\mu \nu} \right) + \cos \theta \, \partial^2 h_{\mu \nu} = - 2 \delta \left( l - R \, \bt \right) \frac{\cos \bt}{M^4} \left( T_{\mu\nu}-\frac{T}{3}\eta_{\mu\nu}+M^{4}
\left[\hat{\zeta}_{,\mu \nu}\right]_{J} \right)
\label{eqall}
\end{equation}

To solve this equation, we construct a (retarded) Green function from the bulk wave functions of the tensor mode, which we then truncate to the zero mode. This truncation is sufficient to describe gravity at large distances,
 which is our purpose. The computation is presented in appendix~\ref{appG}.  Analogously to what found in~\cite{gartan}, the zero--mode truncated solution of~(\ref{eqall}) is
\begin{eqnarray}
h_{\mu \nu}^{(0)} &\equiv& h_{\mu \nu}^{(m)} + h^{(\zeta)}_{,\mu\nu} \label{hsplit} \\
h_{\mu\nu}^{(m)} &\equiv& - \frac{4 \pi R \beta}{M^4 \, V_2} \cos \bt \: 
\left(\partial^2\right)^{-1}
\left[ T_{\mu\nu}- \frac{T}{3} \eta_{\mu\nu} \right] \label{hm} \\
h^{(\zeta)} &\equiv& - \frac{4 \pi R \beta}{V_2} \cos \bt \: \left(\partial^2\right)^{-1}
\left[ {\hat \zeta} \right]_J \label{hzeta}
\end{eqnarray}
where $V_2$ is the volume of half bulk, explicitly given in eq.~(\ref{v2}).

\bigskip

We are interested in the gravitational interaction between zero mode particles localized on the string. The wave function of these particles are evenly spread along the $\phi$ direction and probe the induced metric
\begin{equation}\label{g4}
g^{(4)}_{\mu\nu} = \eta_{\mu \nu} \left( 1 + 2 \Psi\right) + h_{\mu \nu}^{(m)} + \left( h^{(\zeta)} + 2 E \right)_{, \mu \nu}
\end{equation}
where we have truncated to the zero mode~(\ref{hsplit}). We want to compute the linearized Ricci tensor associated to the metric~(\ref{g4}). In general, for a metric $\eta_{\mu \nu} + \delta g_{\mu \nu}$ the linearized Ricci tensor is (the indexes are raised with the flat metric
$\eta^{\mu\nu}$)
\begin{equation}
R^{(4)}_{\mu \nu} = \frac{1}{2} \left( \delta g_{\mu \lambda} \,^{,\lambda}_{,\nu} + \delta g_{\lambda \nu} \,^{,\lambda}_{,\mu} - \delta g_{,\mu \nu} - \partial^2 \delta g_{\mu \nu} \right)
\label{linric}
\end{equation}
It is immediate to verify that the last term in~(\ref{g4}) does not contribute to~(\ref{linric}). This shows that the residual  gauge freedom $\xi \left( x^\mu \right) \,$, mentioned in the discussion after eq.~(\ref{residual}), can be used to set $h^{(\zeta)} + 2 E = 0$ on the string. Then, taking into account that $h_{\mu \nu}^{(0)} = h_{\mu \nu}^{(m)} + h^{(\zeta)}_{,\mu\nu}$ is transverse and traceless, simple algebra gives 
\begin{eqnarray}
R^{(4)}_{\mu \nu} &=& - \frac{1}{2} \partial^2 h_{\mu \nu}^{(m)} + \frac{1}{4} \partial^2 \, \partial^2 \, h^{(\zeta)} \eta_{\mu \nu} +
\left( \frac{1}{2} \partial^2 \Upsilon \eta_{\mu \nu} + \Upsilon_{,\mu\nu} \right) \nonumber\\
\label{ricci}
\end{eqnarray}
where we have introduced the combination
\begin{equation}
\Upsilon \equiv \left( -2\,\Psi - \frac{1}{2} \partial^2 h^{(\zeta)} \right) \Big\vert_{m = \bt}
\end{equation}

From~(\ref{hm}), the first term in~(\ref{ricci}) gives
\begin{equation}
- \frac{1}{2} \partial^2 h_{\mu \nu}^{(m)} = \frac{2 \pi R \beta}{M^4 V_2} \left( T_{\mu \nu} - \frac{T}{3} \eta_{\mu \nu} \right)
\end{equation}
Using~(\ref{trace}), the second term can be related to the deformation mode of the string; using then
eq.~(\ref{hzeta}) we get
\begin{equation}
\frac{1}{4} \partial^2 \partial^2 h^{(\zeta)} \eta_{\mu \nu} = - \frac{\pi R \beta}{V_2} \left[ \partial^2 {\hat \zeta} \right]_J =
\frac{2 \pi R \beta}{M^4 V_2} \left( - \frac{T}{6} \eta_{\mu \nu} \right)
\end{equation}

Therefore,
\begin{eqnarray}
R^{(4)}_{\mu \nu} &=& \frac{2 \pi R \beta}{M^4 V_2} \cos \bt \left[ T_{\mu \nu} - \frac{T}{3} \eta_{\mu \nu} - \frac{T}{6} \eta_{\mu \nu} \right] + \left( \frac{1}{2} \partial^2 \Upsilon\, \eta_{\mu \nu} + \Upsilon_{,\mu\nu} \right) =\\
\, \!\!\!\!\!\!\!\!\!\!\!\!\!\!\! &=&  \frac{2 \pi R \beta}{M^4 V_2} \cos \bt \left[ T_{\mu \nu} - \frac{T}{2} \eta_{\mu \nu} \right]
+ \left( \frac{1}{2} \partial^2 \Upsilon \,\eta_{\mu \nu} + \Upsilon_{,\mu\nu} \right) \nonumber
\end{eqnarray}
Precisely as in~\cite{gartan}, the contribution of the deformation mode ``changes'' the $T/3$ factor into the standard four dimensional one $T/2$.

The energy--momentum tensor $T_{\mu \nu}$ represents matter on the five dimensional string (indeed, it has mass dimension $5$). For the matter zero modes we are considering, the four dimensional tensor is obtained by integrating along the (homogeneous) $\phi$ direction
\begin{equation}
T_{\mu \nu}^{(4)} = \int_0^{2 \pi R  \beta} d \phi \sqrt{\gamma_{\phi \phi}^{(0)}} T_{\mu \nu} = 2 \pi R \beta \cos{\bt}\, T_{\mu \nu}
\end{equation}
With this into account,
\begin{equation}
R^{(4)}_{\mu \nu} = \frac{1}{M^4 V_2} \left[ T_{\mu \nu}^{(4)} - \frac{T^{(4)}}{2} \eta_{\mu \nu} \right]  + \left( \frac{1}{2}\, \partial^2 \Upsilon \, \eta_{\mu \nu} + \Upsilon_{,\mu\nu} \right)
\label{final}
\end{equation}
In absence of the second term, we would then recover ordinary gravity at large distance, in terms of the $4$d Planck mass
\begin{equation}
M_p^2 \equiv M^4 \, V_2
\end{equation}
which is exactly the general ADD relation for flat extra dimensions~\cite{add}. 

The term in $\Upsilon$ signals a scalar--tensor theory of gravity. Indeed, if we define a new $4$--dimensional metric 
\begin{equation}\label{ein}
\bar{g}^{(4)}_{\mu\nu}\equiv \left(1+\Upsilon\right)\,g^{(4)}_{\mu\nu}\,\,,
\end{equation}
the corresponding Ricci tensor shall obey the equation
\begin{equation}
\bar{R}^{(4)}_{\mu \nu} = \frac{1}{M^4 V_2} \left[ T_{\mu \nu}^{(4)} - \frac{T^{(4)}}{2} \eta_{\mu \nu} \right]\,\,.
\end{equation}
This transformation brings the system to the Einstein frame, where (at least the linearized level) the action of matter is of the standard form, but in terms of the metric $\bar{g}^{(4)}_{\mu\nu}$.

As we now show, the effect of $\Upsilon$ is in any case negligible at large distances (that is, at distances 
much larger than the compactification scale). By using~(\ref{hzeta}), we can rewrite
\begin{equation}
\Upsilon = -2\,\Psi + \frac{2 \pi R \beta}{V_2} \cos \bt \left[ {\hat \zeta} \right]_J
\end{equation}
and, by inserting the expressions~(\ref{psisol}) for $\Psi \,$,
and~(\ref{zeta-out}) and~(\ref{zeta-in}) for ${\hat \zeta} \,$, one explicitly finds
\begin{eqnarray}
\Upsilon &=& D_\Psi^{(i)} \frac{1 - \sin\bt}{3\,\left[ \sin\bt + \beta\, \left( 1 - \sin\bt \right) \right]} \, \left[ 9\,\beta + \left( 10 - 7 \, \beta \right) \, \sin\bt- \left( 11 + \beta \right) \, \sin^2 \bt \, \left( 1 + \sin\bt \right) \right] \nonumber\\
&\equiv& D_\Psi^{(i)}\,\frac{1-\sin\bt}{\sin\bt+\beta\,\left(1-\sin\bt\right)}\,F\left(\beta,\,\bt\right)
\label{upsilon}
\end{eqnarray}
The most important information that we obtain from this relation is that only the mode
$D_\Psi^{(i)}\,$ enters in $\Upsilon$. As shown in Appendix~\ref{appH}, also the perturbations of the volume of the internal space is proportional only to $D_\Psi^{(i)}\,$ (the same is true also for the circumference of the cylinder). Not surprisingly, we see that exact standard $4$d gravity is recovered only when the gravitational interaction does not perturb the internal space.

The mode $D_\Psi^{(i)}\,$ can be related to the energy--momentum tensor of matter through eq.~(\ref{psijtt}). This gives
\begin{equation}
D_\Psi^{(i)}=-\frac{\cos\bt}{4}\,\frac{R\,\beta}{M^4}\,\left(\frac{T}{3}-T^\phi_\phi\right)\,\,.
\label{eqd}
\end{equation}
With this taken into account, eq.~(\ref{final}) rewrites
\begin{equation}
R^{(4)}_{\mu\nu}=\frac{1}{M_p^2}\,\left[T_{\mu\nu}^{(4)}-\frac{T^{(4)}}{2}\,\eta_{\mu\nu}\right]-\frac{R^2\,\beta}{4\,M_p^2}\,\left(1-\sin\bt\right)\,F\left(\beta,\,\bt\right)\,\,\left(\frac{1}{2}\,\eta_{\mu\nu}\der^2+\der_\mu\der_\nu\right)\,\left(\frac{T^{(4)}}{3}-T^{(4) \, \phi}_{\;\;\;\; \phi} \right)
\label{final2}
\end{equation}
It is now immediate to see that the second term is negligible at distances much larger than the compactification scale, for which the operator $R^2 \, \partial^2$ is much smaller than one. Moreover, we see that this term decouples as the string shrinks to the pole ($\bt \rightarrow \pi / 2$), since $F \left( \beta ,\, \bt \right)$ is regular in this limit. This concludes our proof that gravity at large distance in this system is simply Einstein $4$d gravity. We will comment further on the stabilization of the internal space and on the role of $\Upsilon$ in the following concluding section.

\section{Conclusions and Outlook}

In this paper 
we considered a brane-world model in a six dimensional space-time, in which
two of the dimensions are compactified by a flux~\cite{ss}. 
The brane-world can be seen as a thick codimension two brane, embedded in the six dimensional space and coupled to the bulk gauge field.

Our study addresses  the observation that  only pure tension can be localized on a strict codimension two defect embedded in six dimensions~\cite{cline,gregory}. 
To overcome this problem, 
we replaced the codimension two brane by a cylinder, centered on the axis of symmetry of the system. 
  We showed that fields with arbitrary energy--momentum tensor can be placed on the cylinder, while the bulk geometry remains regular everywhere. More importantly, the 
gravitational interaction between two zero modes on the string is $4$d Einstein gravity at distances $L$ much greater than the compactification radius $R$ of the internal space.

To see this, we have studied the massless perturbations of this geometry. After solving the complete set of equations, only three modes are relevant for the gravitational interaction; a tensor and two scalar ones. 
One of the two scalars plays the same role as the brane bending mode in the RS model~\cite{gartan}. 
Its presence modifies the Einstein equation so to reproduce the exact tensorial structure of the $4$d Einstein tensor (calling $R^{(4)}$ the 4 dimensional Ricci scalar, 
it modifies the $-R^{(4)}/3$ coefficient, typical of codimension one, into the standard $-R^{(4)}/2$). In our case the bending is associated to (azimuthal symmetric) small deformations of the radius of the string which occur where matter is localized. The second mode, denoted by $D_\Psi^{(i)}$ in the paper, instead modifies the Einstein equations, so that gravity in this model is actually of the scalar--tensor type. 
However, we showed that this modification is parametrically suppressed by the ratio $\left( R / L \right)^2 \,$. Therefore, gravity at large distances is the standard $4$d Einstein one.

Of the two scalars, it is interesting to notice that only the mode $D_\Psi^{(i)}$ controls the volume of the internal space. Therefore, the correction to $4$d gravity is easily seen to be associated to a deformation of the compact dimensions. This situation is analogous to the one emerged in the cosmological studies of codimension one brane-worlds, where it was found that the stability of the internal space is mandatory to recover standard $4$d cosmology~\cite{kanti,cgrt}. As in that case, the mode is coupled to the trace of the stress-energy tensor of the brane fields. So, unless a brane field has a traceless energy--momentum tensor (which can be arranged for by providing a suitable pressure along the compact angular dimension) it will deform the volume of the internal space, resulting in a scalar--tensor gravitational interaction. The smallness of this effect at large scales can be attributed to the fact that 
both the volume of the internal space and the brane position are stabilized
 by fluxes in absence of matter on the string (this can be attribute to flux conservation~\cite{gapo} in general, and also to quantization conditions if charged fields are present in the bulk). 
 
The contribution due to $D_\Psi^{(i)}$ is not the only effect that we expect at small scales. At distances comparable or smaller than the compactification radius, the KK modes of the perturbations of the geometry will become relevant. In general, we expect two sources of corrections to the $4$d theory at short scale. The ones from gravity become effective at distances comparable to the compactification radius $R$. However, a second class of modifications emerges from the Standard Model interactions between the KK modes of the fields living on the string. The latter manifest themselves at distances comparable to the radius of the string, $R \cos \bt \,$. Unless the string is hierarchically close to the pole, 
the effects from the Standard Model interactions will dominate. In principle, one can even envisage the rather peculiar situation where both effects are of phenomenological interest (for instance, a string of length $\sim \left( 10\, {\rm TeV} \right)^{-1}$ in a sub--millimeter bulk).

A relevant question for our construction is what happens in the limit where the string is shrunk to a codimension two brane at the pole, $\bt \rightarrow \pi / 2 \,$. The correct procedure is to keep $T_{\mu \nu}^{(4)}$ constant in this limit ($T_{\mu \nu}^{(4)}$ is the matter energy--momentum tensor integrated along the length of the string, and it is the quantity entering in the $4$d equations). As eq.~(\ref{trace}) then shows, the perturbation of the brane position, denoted by ${\hat \zeta}$ in the paper, diverges in this limit. As soon as ${\hat \zeta}$ exceeds the unperturbed radius of the string, the perturbative approximation breaks down, and the behavior of gravity cannot be anymore obtained by our computation. Given the difficulties with the strict codimension two case, we argue that the gravitational interaction mediated by the scalar modes becomes strong in this limit.
A further complication for the comparison between the two cases is the fact that scalar modes are absent if the brane has codimension $2 \,$ (see the first of ~\cite{precedenti2}). An intuitive understanding of this fact can be gained by noting that the deformation of the string radius does not have a counterpart when the brane is point--like; this is actually the reason which prompted us to introduce the codimension one string, to have a necessary ``brane bending'' mode which would be otherwise absent in codimension two. It is also instructive to compare this discussion with the findings of~\cite{kaloperBH}, where a relativistic shockwave was obtained in the codimension two case. The mode ${\hat \zeta}$ is  actually coupled to the trace of $T_{\mu \nu}^{(4)} \,$. Therefore, the bending mode is not excited by a relativistic source; we can then expect that gravity remains weak in this case, and the limit $\bt \rightarrow \pi / 2$ is continuous, in agreement with what argued in~\cite{kaloperBH}.

\smallskip

There are several open issue which are worth further analysis. One is the actual stability of the compactification we have considered. We studied the massless modes of the theory, but we did not exclude the possibility that tachyonic modes are present (particularly, in the scalar sector). Another interesting issue is the cosmology of the model, when the matter fields on the string give a sizable contribution to the background evolution. In the case we have studied, the position of the string is governed, among other thing, by the tension on it. It is possible that a time dependent energy density would result in a motion of the string in the internal space, while the non-compact coordinates are expanding. A non perturbative analysis on the lines of the one in~\cite{lpps} may be able to clarify this aspect. 
Finally, let us comment that a construction like ours can be straightforwardly
performed in a  supergravity framework, by considering Salam-Sezgin
compactification of Nishino-Sezgin gauged supergravity \cite{nishino}, by 
including also the perturbations of the scalar and three-form fluxes
that appear on that system.

\vspace{0.9cm}
\centerline{\bf Acknowledgments}
\vspace{0.1cm}

\noindent
We thank Cliff Burgess, Emilian Dudas, Nemanja Kaloper, and Ignacio Navarro for very useful discussions. The work of M.P. was supported in part by the Department of Energy under contract
DE-FG02-94ER40823, and by a grant from the Office of the Dean of the Graduate School of the University of Minnesota. G.~T.~is partially supported by the EC $6^{th}$ Framework Programme MRTN-CT-2004-503369.

\appendix
\def\theequation{\thesection.\arabic{equation}}
\setcounter{equation}{0}

\begin{center}
\section*{Appendixes}
\end{center}

\bigskip

\appendix
\def\theequation{\thesection.\arabic{equation}}
\setcounter{equation}{0}

\section{Flux quantization}
\label{appA}

In this appendix we review how, in presence of a bulk fermion charged under the $U \left( 1 \right)$ symmetry, the background solution is quantized.  The gauge field leading to eq.~(\ref{intrgf}) has to be taken as
\begin{eqnarray}
A_\phi &=& M^2 \left( \sin \theta -1 \right) \equiv A_\phi^{(N)} \quad\quad {\rm for}
\;   l \ge - l_{D}  \\
A_\phi &=& M^2 \left( \sin \theta + 1 \right) \equiv A_\phi^{(S)} \quad\quad {\rm for}
\; - l \le l_{D}
\,\,.
\end{eqnarray}
where $l_{D} > 0$. This guarantees that $A_\phi$ is well defined ($A_\phi = 0$) at the two poles, and that we have a region $- l_D <0< l_D$ where $A_\phi$ is doubly defined:
\begin{equation}
A_\phi^{(N)} - A_\phi^{(S)}= -  2\,M^2 \,\,,
\end{equation}
However, the difference between the two values is unphysical if it corresponds to the gauge transformation
\begin{equation}
A_\phi \rightarrow A_\phi + \der_\phi \Lambda \;\;\;,\;\;\; \Lambda = -2\,
M^2 \phi \,\,.
\label{gaugetransf}
\end{equation}
The quantization arises in presence of a bulk field which is charged the $U \left( 1 \right)$ symmetry. In correspondence of~(\ref{gaugetransf}), a bulk fermion of charge $\e$ transforms as
\begin{equation}
\psi \rightarrow \e^{-i \e \Lambda} \psi = \e^{ 2 i \e M^2 \phi} \psi \,\,.
\end{equation}
The single valuedness for $\psi$ as we send $\phi \rightarrow \phi + 2 \pi \beta R$
imposes the quantization condition~(\ref{quantization}).

\section{Gauge invariant perturbations and bulk equations}
\label{appB}

Under the infinitesimal coordinate transformation $x^A \rightarrow x^A + \xi^A \left( l ,\, x^\mu \right) \,$, with $\xi^\mu = \partial^\mu \xi + {\hat \xi}^\mu \,$ (${\hat \xi}^\mu$ transverse), the perturbations of the system, eqs.~(\ref{perturbedds})and~(\ref{pertgauge}) transform as
\begin{eqnarray}
{\rm tensor}: && h_{\mu \nu} \rightarrow h_{\mu \nu} \nonumber\\
{\rm vectors}: && T_\mu \rightarrow T_\mu - {\hat \xi}_\mu' \nonumber\\
&& V_\mu \rightarrow V_\mu \nonumber\\
&& E_\mu \rightarrow E_\mu - {\hat \xi}_\mu \nonumber\\
{\rm scalars}: && \Phi \rightarrow \Phi - \xi^{l '} \nonumber\\
&& A \rightarrow A - \cos^2 \theta \, \xi^{\phi '} \nonumber\\
&& C \rightarrow C + \theta'  \, \tan \theta \, \xi^l \nonumber\\
&& T \rightarrow T - \xi^l - \xi' \nonumber\\
&& V \rightarrow V - \cos^2 \theta \, \xi^{\phi } \nonumber\\
&& E \rightarrow E - \xi \nonumber\\
&& \Psi \rightarrow \Psi \nonumber\\
{\rm gauge}: && a_l \rightarrow a_l - M^2 \left(\sin \theta - 1 \right)  \, \xi^{\phi '} \nonumber\\
&& a_{\phi} \rightarrow a_{\phi} - \theta' M^{2} \cos{\theta} \xi^{l} \nonumber\\
&& a \rightarrow a - M^2 \left( \sin \theta - 1 \right) \xi^\phi \nonumber\\
&& {\hat a}_\mu \rightarrow {\hat a}_\mu
\label{transf}
\end{eqnarray}
we can now explicitly verify that it is possible to set $E_\mu = T = V = 0 \,$, as done in the main text, with the  residual freedom~(\ref{residual}). It is also immediate to verify that the combinations~(\ref{invariant1}) and (\ref{invariant2}) do not change under~(\ref{residual}).

In the bulk, we have the Maxwell and Einstein equations
\begin{eqnarray}
&& \partial_M \left( \sqrt{- g} g^{MA} g^{NB} F_{AB} \right) = 0 \nonumber\\
&& G_{AB} = \frac{T_{AB}}{M^4}
\end{eqnarray}
which we linearize at first order in the perturbations.

Let us start from the Maxwell equations. The $l ,\, \phi, \,$ and $\mu$ components read, respectively,
\begin{eqnarray}
&&\der^\mu f_{\mu l} = 0 \,\,, \label{maxth} \\
&&\theta' \, \tg \theta \, f_{l \phi} - M^2 \theta' \cos \theta \left( C' + \Phi' - 4 \Psi' - \partial^2 E' \right)  + f_{l \phi}' + \partial^\mu f_{\mu \phi} = 0 \,\,, \label{maxphi} \\
&&\theta' \tg \theta \, f_{\mu l} - f_{\mu l}' - \der^\nu f_{\mu \nu} - \frac{M^2 \theta'}{\cos \theta} V_\mu' = 0 \label{maxmu}
\end{eqnarray}

We want to rewrite these equations in terms of the perturbations of the gauge field, eq.~({\ref{pertgauge}). To do this properly, we must take into account the $U \left( 1 \right)$ gauge symmetry in the bulk. Under an infinitesimal gauge transformations, $\delta A_A \rightarrow \delta A_A + \partial_A \lambda \,$, the different components transform as
\begin{equation}\label{u1gauge}
a_l \rightarrow a_l + \lambda' \;\;\;,\;\;\;
a_\phi \rightarrow a_\phi \,\,,
\end{equation}
where we have assumed that $\lambda$ does not depend on $\phi$.
Since we have decomposed $a_\mu\,$ in a traceless and a transverse part, we also decompose $\lambda = \lambda_H + \lambda_{NH} \,$, where $\partial^2 \lambda_H = 0 \,$. Then,
\begin{eqnarray}
&& a_\mu \rightarrow a_\mu + \der_\mu \left( \lambda_H + \lambda_{NH} \right) \;\;\; \Rightarrow \nonumber\\
&& a \rightarrow a + \lambda_{NH} \;\;,\;\; {\hat a}_\mu \rightarrow {\hat a}_\mu + \der_\mu \lambda_{H}
\end{eqnarray}
(in this way, ${\hat a}_\mu$ remains transverse). Accordingly, the component $f_{\mu l}$ of the field strength is decomposed in
\begin{equation}
f_{\mu l} = \der_\mu \left( a_l - a' \right) - {\hat a}_\mu' \equiv f_{\mu l}^S + f_{\mu l}^V \,\,,
\end{equation}
The scalar and vector components of $f_{\mu l}$ are not separately gauge invariant (although their sum obviously is). Under the gauge transformation, we have
\begin{equation}
f_{\mu l}^S \rightarrow f_{\mu l}^S + \partial_\mu \lambda_H'
\label{u1fs}
\end{equation}
so we can use $\lambda_H$ to set $f_{\mu l}^S = 0$ in the bulk. Note that this does not fix uniquely $\lambda_H$, that still depends on an arbitrary function  $\lambda_H^{\mathrm {res}}\left(x^\mu\right)$. 
Eq.~(\ref{maxth}) is automatically satisfied, since ${\hat a}_\mu$ is transverse, while eq.~(\ref{maxmu}) has only the nontrivial vector component
\begin{equation}
{\hat a}_\mu'' - \theta' \tan \theta \, {\hat a}_\mu' + \partial^2 {\hat a}_\mu - \frac{M^2 \theta'}{\cos \theta} V_\mu' = 0
\label{mxwv}
\end{equation}

\smallskip
Eq.~(\ref{maxphi}) has instead only the scalar component. We can rewrite it in terms of gauge invariant fields defined in the main body of the paper
\begin{equation}
\partial^2 {\hat a}_\phi + {\hat a}_\phi '' + \theta' \tan \theta \, {\hat a}_\phi' - M^2 \theta' \cos \theta \left[ {\hat C}' + {\hat \Phi}' - 4 \Psi' \right] = 0
\label{mxws}
\end{equation}

\smallskip

We can now turn to the Einstein equations for the perturbations. As for the Maxwell equations, we decompose them into tensor / vector / scalar components, and we then write them in terms of gauge invariant fields. For the tensor mode, we obtain
\begin{equation}
\partial^2 h_{\mu \nu} + h_{\mu \nu}'' - \theta' \tan \theta \, h_{\mu \nu}' = 0
\label{eint}
\end{equation}
from the $(\mu \nu)$ equation. For the vector modes, we have instead
\begin{eqnarray}
&& \!\!\!\!\!\!\!\!\!\!\!\!\!\!\!\partial^2 T_\mu = 0 \label{einv1} \\
&& \!\!\!\!\!\!\!\!\!\!\!\!\!\!\!\partial^2
 V_\mu + V_\mu '' + \theta' \tan \theta \, V_\mu ' + \frac{2 \cos \theta \, \theta' {\hat a}_\mu'}{M^2} = 0 \label{einv2} \\
&& \!\!\!\!\!\!\!\!\!\!\!\!\!\!\!T_\mu ' - \theta' \tan \theta \, T_\mu  = 0 \label{einv3}
\end{eqnarray}
from the $(l \mu) \,$, $(\phi \mu) \,$, and $(\mu \nu)$ components, respectively. Finally, in the scalar sector we find

\begin{eqnarray}
&& \theta' \left( {\hat C} + {\hat \Phi} \right) - 4 \tan \theta \, \Psi' + \frac{1}{\theta'} \partial^2 \left( {\hat C} + 3 \Psi \right) =
 \frac{{\hat a}_\phi'}{M^2 \cos \theta} \label{eins1} \\
&& \partial^2 A = 0 \label{eins2} \\
&& 4 \Psi'' + \partial^2 \left( {\hat \Phi} + 3 \Psi \right) + \theta^{' 2} \left( {\hat C} + {\hat \Phi} \right)
= \frac{\theta' {\hat a}_\phi'}{M^2 \cos \theta} \label{eins3} \\
&&   \theta'\, \tan \theta \, \left( {\hat C} - {\hat \Phi} \right)-3 \Psi'
- {\hat C}' = \frac{\theta' \, {\hat a}_\phi}{M^2 \, \cos \theta} \label{eins4} \\
&& A' - \theta' \tg \theta \, A =0 \label{eins5} \\
&&\partial^2 \left( {\hat C} + {\hat \Phi} + 2 \Psi \right) + \theta' \tan \theta \left( - 2 {\hat C}' + {\hat \Phi'} - 3 \Psi' \right) + \nonumber\\
&& +  {\hat C}'' + 3 \Psi'' + \theta^{' 2} \left( {\hat \Phi} - {\hat C} \right) + \frac{\theta' {\hat a}_\phi'}{M^2 \cos \theta} = 0 \label{eins6} \\
&& {\hat C} + {\hat \Phi} + 2 \Psi = 0 \label{eins7}
\end{eqnarray}
from the $(l l) \,$, $(l \phi) \,$, $(\phi \phi) \,$, $(l \mu) \,$, $(\phi \mu) \,$, diagonal $(\mu \nu) \,$, and non-diagonal $(\mu \nu) \,$ components, respectively.

\section{Bulk solutions for the zero modes}
\label{appC}

In this Section, we solve the equations for the zero modes ($\partial^2 \equiv 0$) in the bulk, working out the three sectors separately. The restriction to the zero modes is because we are interested in gravity at much larger distances then the inverse KK masses. In many steps we rename integration constants without mentioning it explicitly.

\smallskip
{\bf Tensor modes}: Eq.~(\ref{eint}) is solved by
\begin{equation}
h_{\mu \nu} = C_{\mu \nu}^{(1)} \left( x \right) + C_{\mu \nu}^{(2)} \left( x \right) \int \frac{d l}{\cos \theta}
\end{equation}
The second term vanishes on the outside, by parity considerations (it would be odd at the equator), and inside, since it would diverge at the pole. Therefore,
\begin{eqnarray}
h_{\mu \nu} = \left\{
\begin{array}{l}
C_{\mu \nu}^{(o)} \left( x \right) \;\;\;,\;\;\; {\rm outside} \\
C_{\mu \nu}^{(i)} \left( x \right) \;\;\;,\;\;\; {\rm inside}
\end{array}
\right.
\end{eqnarray}

\smallskip
{\bf Vector modes}: Eqs.~(\ref{mxwv}) and (\ref{einv2}) can be solved to obtain the zero modes of ${\hat a}_\mu$ and of $V_\mu$ in the bulk. However, these modes do not couple to matter on the string (they do not enter in the junction conditions), and we can simply ignore ignore them.

For the remaining vector $T_\mu \,$, eq.~(\ref{einv1}) indicates that only the zero mode is present, while
eq.~(\ref{einv3}) is solved to give $T_\mu = \tau_\mu \left( x \right) / \cos \theta \,$. The integration ``constant'' $\tau_\mu$ must vanish both on the outside bulk -- since $T_\mu$ should be odd across the equator -- and on the inside bulk -- since otherwise the mode would diverge at the pole. For this reason,
\begin{equation}
T_\mu = 0 \;\;\;,\;\;\; {\rm everywhere}
\end{equation}

\smallskip
{\bf Scalar modes}: We start by rearranging and individuating the set of independent equations among the Einstein / Maxwell ones. Clearly, equations~(\ref{eins2}) and~(\ref{eins5}) are independent. The first one indicates that there are no KK modes, while the second one that the zero mode is $A = D \left( x \right) / \cos \theta \,$. This mode vanishes outside, by parity considerations (it should be odd), and inside, since it would diverge at the pole. Therefore,
\begin{equation}
A = 0
\end{equation}

Among the other ones, eq.~(\ref{mxws}) can be obtained from~(\ref{eins1}) and~(\ref{eins3}), while eq.~(\ref{eins6}) from eq.~(\ref{eins4}). We can simplify the remaining equations by combining eq.~(\ref{eins1}) and eq.~(\ref{eins3}) to obtain
\begin{equation}
\Psi'' + \theta' \tan \theta \Psi' = 0
\label{eins8}
\end{equation}
We then use eq.~(\ref{eins7}) to simplify eqs.~(\ref{eins1}) and(\ref{eins4}),
\begin{eqnarray}
&& \!\!\!\!\!\!\!\!\!\! 2 \theta' \Psi + 4 \tan \theta \, \Psi' + \frac{{\hat a}_\phi'}{M^2 \cos \theta} = 0 \label{eins9} \\
&& \!\!\!\!\!\!\!\!\!\! 2 \theta' \tan \theta \left( {\hat C} + \Psi \right) - 3 \Psi' - {\hat C}' = \frac{\theta' \, {\hat a}_\phi}{M^2 \, \cos \theta} \label{eins10}
\end{eqnarray}
Eqs.~(\ref{eins8}), (\ref{eins9}), (\ref{eins10}), and~(\ref{eins7}), can be then solved in this order.

From eq.~(\ref{eins8}) we find
\begin{eqnarray}
\Psi = \left\{
\begin{array}{l}
C_\Psi^{(o)} \left( x \right) \;\;\;\;\;\;\;\;\;\;\;\;\;\;\;\;\;\;\;\;\;\;,\;\;\; {\rm outside} \\ \\
C_\Psi^{(i)} \left( x \right) + D_\Psi^{(i)} \, \sin \theta \;\;\;,\;\;\; {\rm inside}
\end{array}
\right.
\label{psi-outin}
\end{eqnarray}
where in the outside part one mode vanishes due to the parity symmetry at the equator. Then, from eq.~(\ref{eins9}) we have
\begin{eqnarray}
{\hat a}_\phi = \left\{
\begin{array}{l}
- 2 M^2 C_\Psi^{(o)} \left( x \right) \sin \theta \;\;\;\;\;\;\;\;\;\;\hskip 1.5cm\;\;\;\;\;\;\;\;\;,\;\;\; {\rm outside} \\ \\
M^2 \left[ 2 C_\Psi^{(i)} \left( 1 - \sin \theta \right) + 3 D_\Psi^{(i)} \cos^2 \theta \right]
\;\;\;\;, \;\;\;{\rm inside}
\end{array}
\right.
\end{eqnarray}
where the integration constant has been fixed, on the outside, by the parity symmetry at the equator, and on the inside, by the fact that ${\hat a}_\phi$ must vanish at the pole. Eq.~(\ref{eins10}) gives now
\begin{eqnarray}
{\hat C} = \left\{
\begin{array}{l}
- 2 C_\Psi^{(o)} \left( x \right) + C_C^{(o)}\,\frac{1}{\cos^2 \theta} \;\;\;\;\;\;\;\;\;\;
\hskip 1.5cm
\;\;\;\;\;\;\;\;\;,\;\;\; {\rm outside} \\ \\
- 2 C_\Psi^{(i)}\,\frac{\sin \theta}{1+ \sin \theta} +
 \frac{2}{3}\, D_\Psi^{(i)} \frac{\left(5 - 4 \sin \theta - 4 \sin^2 \theta \right)}{1 + \sin \theta} \;\;\hskip 0.5cm,\;\;\hskip 0.2cm {\rm inside}
\end{array}
\right.
\end{eqnarray}
where the integration constant inside vanishes due to the request of regularity of ${\hat C}$ at the pole. Finally, ${\hat \Phi}$ simply follows from the algebraic equation~(\ref{eins7}).

\section{Induced metric and continuity conditions}
\label{appD}

We choose the brane to remain at the unperturbed position $\bt \,$. In this case
\begin{equation}
{\hat \zeta} = - E'
\end{equation}
and the induced metric is given in~(\ref{induced}) (vector modes excluded). $E$ can be gauged to zero at the brane; we do not consider it further. We require the continuity of $h_{\mu \nu} ,\, \Psi ,\, C ,\, a_\phi \,$ (as well as of $a$ and $a_l$). The continuity of $h_{\mu \nu}$ gives $C_{\mu \nu}^{(o)} = C_{\mu \nu}^{(i)} \equiv C_{\mu \nu} \,$. So,
\begin{equation}
h_{\mu \nu} = C_{\mu \nu} \left( x \right) \;\;\;,\;\;\; {\rm everywhere}
\label{hsol2}
\end{equation}
The continuity of $\Psi$ gives instead $C_\Psi^{(o)} = C_\Psi^{(i)} + D_\Psi^{(i)} \sin \bt \,$, and so
\begin{eqnarray}
&& \Psi = \left\{
\begin{array}{l}
C_\Psi^{(i)} + D_\Psi^{(i)} \sin \bt  \;\;\;\;\;\;\;\;\;\:\,,\;\;\; {\rm outside} \\ \label{psisol} \\
C_\Psi^{(i)}  + D_\Psi^{(i)} \, \sin \theta \hskip 1cm\;\;\;,\;\;\; {\rm inside}
\end{array}
\right. \\ \nonumber\\
&& {\hat a}_\phi = \left\{
\begin{array}{l}
- 2 M^2 \left[ C_\Psi^{(i)} + D_\Psi^{(i)} \sin \bt \right] \sin \theta
\hskip 1.5cm\;,\;\; {\rm outside} \\ \\
M^2 \left[ 2 C_\Psi^{(i)} \left( 1 - \sin \theta \right) + 3 D_\Psi^{(i)} \cos^2 \theta \right]\;\; ,
\hskip 0.3cm {\rm inside}
\end{array}
\right. \\\nonumber
\end{eqnarray}

The continuity conditions of $C$ and $a_\phi$ rewrite
\begin{eqnarray}
&& \left[ {\hat C} - \theta' {\hat \zeta} \tan \bt \right]_J = 0 \nonumber \\
&& \left[ {\hat a}_\phi + M^2 \theta' {\hat \zeta} \cos \bt \right]_J = 0
\end{eqnarray}
which give
\begin{eqnarray}
C_C &=& 2 C_\Psi^{(i)}  + \frac{D_\Psi^{(i)}}{3} \left( 10 - 3 \sin \bt -  \sin^3 \bt \right) \nonumber\\
\left[ \theta' {\hat \zeta} \right]_J &=& \frac{2 C_\Psi^{(i)} + D_\Psi^{(i)} \left( 3 - \sin^2 \bt \right)}{\cos \bt}
\label{zj}
\end{eqnarray}

For future use, we explicitly write also the mode ${\hat \Phi}$, which evaluates to
\begin{eqnarray}
{\hat \Phi} = \left\{
\begin{array}{l}
\frac{1}{\cos^2 \theta} \left[ - 2 C_\Psi^{(i)} - \frac{D_\Psi^{(i)} \left( 10 -3 \sin \bt - \sin^3 \bt \right)}{3} \right]
, {\rm outside} \\\nonumber\\
- \frac{2 C_\Psi^{(i)}}{1+ \sin \theta} -\frac{2 D_\Psi^{(i)} \left[ 5 - \sin \theta - \sin^2 \theta \right]}{3 \left( 1 + \sin \theta \right)} \;\;,\;\; {\rm inside}
\end{array}
\right.
\label{phi-hat}
\end{eqnarray}

We conclude this appendix by verifying that the perturbed solution is indeed regular at the pole.
We start from the line element
\begin{equation}
d s^2 = R^2 \beta^2 \left( 1 + 2 \Phi \right) d \theta^2 + \left( 1 + 2 C \right) \cos^2 \theta \, d \phi^2
\end{equation}
for the inside bulk, and introduce the radial coordinate
\begin{equation}
r = R \beta \int_\theta^{\pi/2} \left( 1 + \Phi \left( {\tilde \theta} \right) \right) d {\tilde \theta}
\end{equation}
so that $r = 0$ at the pole. In this system of coordinates, an explicit computation gives
\begin{equation}
d s^2 = d r^2 + \cos^2 \theta_0 \left[ 1+ \frac{20 \cos^2 \theta_0}{3 \left( 1 + \sin \theta_0 \right)} D_\Psi^i \right] d \phi^2 \;\;\;,\;\;\; \theta_0 \equiv \frac{\pi}{2} - \frac{r}{R \beta}
\end{equation}
Expanding for $r \rightarrow 0 \,$ (that is, in a neighborhood of the pole), the perturbation piece drops (being of order $r^4$), and we are left with
\begin{equation}
d s^2 \simeq d r^2 + r^2 \, d \left( \frac{\phi}{R \beta} \right)^2
\end{equation}
By comparing with eq.~(\ref{extcoord}), we see that the coordinate $\phi / R \beta$ has period $2 \pi$ (no deficit angle) which confirms that the inside solution is regular.

\section{Maxwell junction conditions}
\label{appE}

Varying the total action with respect to the gauge field gives the Maxwell equations at the brane
\begin{equation}
\left[ \sqrt{-g} F^{l B} \right]_J -\e\, v^2 \,\sqrt{-\gamma}\,\left( \partial^B\sigma - \e A^B\right) = 0
\end{equation}
Lowering the indices, and using $\sqrt{-g} = \sqrt{g_{l l}} \sqrt{-\gamma} \,$, we can write
\begin{equation}\label{mxjc}
\left[ \sqrt{g^{l l}} F_{l B} \right]_J -\e\, v^2 \,\left( \partial_B\sigma - \e A_B\right) = 0\,\,.
\end{equation}
The $l$ component of this equation gives $a_l=0$ at the brane.

\ni
The equation for $\delta\sigma$ reads $\partial^2_\phi\delta\sigma=0$. This is solved by $\delta\sigma=\delta\sigma_0 \left(x^\mu\right)+\delta\sigma_1\left(x^\mu\right)\,\phi$. However, periodicity in $\phi$ of $\sigma$, and the fact that $\delta \sigma$ is infinitesimal, imposes $\delta\sigma_1=0$, so that we are left just with $\delta\sigma=\delta\sigma_0\left(x^\mu\right)$.

\ni
The $\mu$ component of eq.~(\ref{mxjc}) finally gives
\begin{equation}
\left[ a' - a_l \right]_J - \e\, v^2 \left(\delta\sigma_0-\e\,a\right) = 0
\end{equation}
Since the first term is zero due to the gauge choice $f_{\mu l}^S = 0 \,$, we have $\delta\sigma_0=\e\, a$ at the brane, that can be set to zero by using the residual gauge  freedom given by $\lambda_H^{\mathrm {res}}$ (see appendix~\ref{appB}).
\smallskip

\ni
 The $\phi$ component gives a background equation, already accounted for, plus the equation for the scalar modes
\begin{equation}
\left[ M^2 \theta' \left( - \cos \bt {\hat \Phi} - \frac{\theta' \, {\hat \zeta}}{\sin \bt} \right) + {\hat a}_\phi' - \frac{\theta' {\hat a}_\phi}{\tan \bt} \right]_J = 0
\end{equation}
Using eq.~(\ref{eins1}), we can rewrite it as
\begin{equation}
\left[ 4 \tan^2 \bt \, \Psi'  - \theta' \left( \tan \bt \, {\hat C} - \frac{{\hat a}_\phi}{M^2 \cos \bt} - \frac{\theta' {\hat \zeta}}{\cos^2 \bt} \right) \right]_J = 0
\label{mxwjunc}
\end{equation}
This equation, together with the second of~(\ref{zj}), allows to determine ${\hat \zeta}$ both outside and inside in terms of $C_\Psi^{(i)}$ and $D_\Psi^{(i)} \,$,
\begin{eqnarray}\label{zeta-out}
&& \left( \theta' {\hat \zeta} \right)_{\rm out} = 2  \, C_\Psi^{(i)} \, \tan \bt+  D_\Psi^{(i)} \,  \tan \bt\, \frac{10 \left( 1- \beta \right) - \left( 15 - 3 \beta \right) \sin \bt + \left( 11 + \beta \right) \sin^3 \bt}{3 \left( 1 - \beta \right)}  \\
\nonumber\\
&& \left( \theta' {\hat \zeta} \right)_{\rm in} = - \frac{2 \cos \bt \, C_\Psi^{(i)}}{1 + \sin \bt} +
D_\Psi^{(i)} \,  \cos \bt \Bigg\{ \label{zeta-in} 
\frac{( 1 - \beta ) \sin \bt - 9 \left( 1 - \beta \right) - \left( 11 + \beta \right) \sin^2 \bt \left( 1 + \sin \bt \right)}{3 \left( 1 - \beta \right) \left( 1 + \sin \bt \right)} \Bigg\} 
\end{eqnarray}

\section{Boundary conditions at the string}
\label{appF}

In this appendix, we compute the second Israel conditions,
\begin{equation}
\left[ \hat{K}_{A B}\right]_{J}=-\frac{S_{A B}}{M^4}
\label{junctions}
\end{equation}
where $S_{AB}$ is the stress energy momentum of the brane, while $\hat{K}_{M N}= K_{M N}-\gamma_{M N} \, K \,$, $K_{AB}$ denotes the extrinsic curvature (and $K$ is its trace). For generality (and as a cross-check), we do not restrict to Gaussian normal coordinates in this computation, and we show that these equations can be written in terms of gauge invariant quantities only. However, in the rest of the paper we then set $\zeta = \Phi = 0$ at the brane, when we deal with the equations derived here (as a consequence, we have then ${\hat \zeta} = - E' ,\; {\hat \Phi} = E'' \,$ at the brane).

$S_{AB}$ is readily computed from~(\ref{actionbrane}); we also add the energy momentum tensor $T_{AB}$ of matter localized on the brane.
We assume that matter
 can be treated perturbatively together with the other perturbations of the system (this is the standard assumption in brane-world models). In this way, the only non vanishing components of the energy--momentum tensor of the string are
\begin{eqnarray}\label{brbackenten1a}
S_{\mu \nu } &=& -\left[ \lambda_s +\frac{v^2}{2}
\left(\partial_\phi\sigma-\e\,A_{\phi}\right)\,\left(\partial^\phi\sigma-\e\, A^{\phi}\right) \right] \,\gamma_{\mu \nu}+ T_{\mu\nu} \\
S_{\phi \phi} &=&  -\left[ \lambda_s - \frac{v^2}{2}
\left(\partial_\phi\sigma-\e\,A_{\phi}\right)\,\left(\partial^\phi\sigma-\e\, A^{\phi}\right) \right] \,\gamma_{\phi\phi} +T_{\phi \phi}
 \label{brbackenten2a}
\end{eqnarray}
where $\partial_\phi\sigma-\e\,A_{\phi}= -\e\,\left(M^2 \, \sin \bt+a_\phi\right)$ at the string.

The intrinsic metric of the string is $\gamma_{AB} = g_{AB} - n_A n_B \,$, where all these quantities (without restricting to Gaussian normal coordinates) are computed at the string location $\bt + \zeta \,$
The normal of the string is, in components
\begin{equation}
n_l = - 1 - \Phi \;,\; n_\phi = 0 \;,\; n_\mu = \partial_\mu \zeta
\end{equation}
(we note that we have chosen it to point towards the outside bulk). The extrinsic curvature, $K_{AB} \equiv \gamma_A^C \nabla_C n_B \,$, at the two sides of the brane has the components
\begin{eqnarray}
K_{\mu \nu} &=& \zeta_{,\mu \nu} - \eta_{\mu \nu} \Psi' - E'_{,\mu \nu} - \frac{1}{2} h'_{\mu \nu} \\
K_{\phi \phi} &=& \theta' \, \sin \bt \, \cos \bt \left( 1 + 2 C - \Phi \right) - \cos^2 \bt \: C' - \theta^{' 2} \left( {\rm sin }^2 \bt - {\rm cos }^2 \bt \right) \zeta
\end{eqnarray}

Evaluating~(\ref{junctions}) we find, for the background, the two relations~(\ref{fixbr1}) and (\ref{defq2})
  given in the main text. These background relations can be also used to write also $S_{AB}$ (apart from the matter component) as a jump. More precisely, inside $S_{AB}$ we have terms of the form $\lambda_s \times f$, where $f$ is a function of the perturbations which is continuous at the two sides of the brane. We rewrite this term as
\begin{equation}
\lambda_s f = - \frac{M^4}{2} \tan \bt \left[ \theta' \right]_J f = - \left[ \frac{M^4}{2} \tan \bt \, \theta' f \right]_J\,\,.
\end{equation}
This significantly simplifies the perturbed part of the~(\ref{junctions}). We obtain
\begin{eqnarray}
&&\left[ \eta_{\mu \nu} \left( \partial^2 {\hat \zeta} - 3 \Psi' - {\hat C}' + \theta' \tan \bt \, \left( {\hat C} - {\hat \Phi} \right) - \frac{\theta' {\hat a}_\phi}{M^2 \cos \bt} \right) - {\hat \zeta}_{,\mu\nu} + \frac{1}{2} h'_{\mu \nu} \right]_J = \frac{T_{\mu \nu}}{M^4} \label{juncmunu} \\
&&\left[ \partial^2 {\hat \zeta} - 4 \Psi' - \tan \bt \, \theta' \, {\hat C} + \frac{\theta' {\hat a}_\phi}{M^2 \cos \bt} +  
\frac{\theta'^2 {\hat \zeta}}{\cos^2 \bt} \right]_J = \frac{T_{\phi}^{\phi}}{M^4} \label{juncphiphi} \\
&& a_{|_{\rm brane}} = 0 \label{juncmuphi}
\end{eqnarray}
which correspond, respectively, to the $(\mu \nu) ,\, (\phi \phi) ,\,$ and $(\mu \phi)$ components of~(\ref{junctions}).
Eq.~(\ref{juncmunu}) simplifies once combined with~(\ref{eins4}), leading to
\begin{equation}
\left[ \eta_{\mu \nu} \partial^2 {\hat \zeta} - {\hat \zeta}_{,\mu\nu} + \frac{1}{2} h'_{\mu \nu} \right]_J = \frac{T_{\mu \nu}}{M^4}
\label{j2-munu}
\end{equation}
Tracing this equation gives the relation~(\ref{trace}) given in the main text. Then, combining~(\ref{j2-munu}) and~(\ref{trace}) we obtain the junction condition~(\ref{j-munu}). Also Eq.~(\ref{juncphiphi}) simplifies once combined with~(\ref{mxwjunc}), leading to the junction condition~(\ref{j-phiphi}).

\section{Putting together bulk and brane}
\label{appG}

The aim of this appendix is to combine the relevant bulk and brane equations, and to provide a full solution for the zero modes in presence of matter on the string. The computations in this and of the next appendix follows the analogous ones of~\cite{gartan}, although we prefer to clarify several intermediate steps. The bulk eq.~(\ref{eint}) for the tensor mode can be combined with the boundary condition~(\ref{j-munu}) to give
\begin{equation}
{\cal O} h_{\mu \nu} \equiv \left[ \partial_l \left( \cos \theta \, \partial_l \right) + \cos \theta \, \partial^2 \right] h_{\mu \nu} =
 - 2\delta \left( l - R \bt \right) \frac{\cos \bt}{M^4} \Sigma_{\mu \nu}
\label{bulkbrane}
\end{equation}
where, analogously to~\cite{gartan}, we have defined
\begin{equation}
\Sigma_{\mu \nu} \equiv T_{\mu\nu}-\frac{T}{3}\eta_{\mu\nu}+M^{4}
\left[\hat{\zeta}_{,\mu \nu}\right]_{J}
\end{equation}

To solve eq~(\ref{bulkbrane}), we construct the (retarded) Green function of
  the  operator ${\cal O} \,$, satisfying
\begin{equation}
{\cal O} \, G \left( x, l; x', l' \right) = \delta^{(4)} \left( x - x' \right) \delta \left( l - l' \right)
\end{equation}
One can verify that it is given by
\begin{equation}
G \left( x , l ; x', l' \right) = - \int \frac{d^4 k}{\left( 2 \pi \right)^4} \, {\rm e}^{i k_\mu \left( x^\mu - x^{'\mu} \right)} \sum_n \frac{\psi_n \left( l \right) \psi_n \left( l' \right)}{p_n^2 + {\bf k}^2 - \left( \omega + i \epsilon \right)^2}
\label{green}
\end{equation}
provided the functions $\psi_n \left( l \right)$ are a complete set of 
eigenfunctions  of
\begin{equation}
\partial_l \left( \cos \theta \, \partial_l \right) \psi_n = - \cos \theta \, p_n^2 \, \psi_n
\label{eigen}
\end{equation}
normalized as~\footnote{The integral is performed for $l > 0 \,$. We could equivalently work with the entire bulk, under the assumption of $Z_2$ symmetry. We would then have to add an equal amount of matter on the string in the other hemisphere.}
\begin{equation}
\int d l \, \cos \theta \, \psi_m \left( l \right) \psi_n \left( l \right) = \delta_{mn}
\label{norma}
\end{equation}
Clearly, the solutions of~(\ref{eigen}) are the bulk wave functions of the tensor modes, where $p_n^2$ is the mass of the mode.

The zero mode truncation amounts in including only the zero mode $\psi_0$ in the sum~(\ref{green}). We already know that $\psi_0$ is a constant, cf. eq.~(\ref{hsol2}). The normalization condition~(\ref{norma}) is
\begin{equation}
1 = \psi_0^2 \int d l \cos \theta =  \frac{\psi_0^2}{2 \pi R \beta} \int_{\rm 1/2-sphere} d l d \phi \sqrt{g_{l l}^{(0)} \, g_{\phi \phi}^{(0)}} = \frac{\psi_0^2}{2 \pi R \beta} \, V_2 
\end{equation}
where $V_2$ is the (background) volume of the bulk, given by
\begin{equation}
V_2 = 2 \pi R^{2}\,\beta\,\left[ \sin{\bar{\theta}}+\beta\left( 1-\sin{\bar{\theta}} \right) \right]
\label{v2}
\end{equation}
Therefore we obtain the zero mode truncated Green function
\begin{equation}
G_0 = - \int \frac{d^4 k}{\left( 2 \pi \right)^4} \, {\rm e}^{i k_\mu \left( x^\mu - x^{'\mu} \right)} \frac{\psi_0^2}{
 {\bf k}^2 - \left( \omega + i \epsilon \right)^2} = \frac{2 \pi R \beta}{V_2} \left(\partial^2\right)^{-1} \delta^{(4)} \left( x - x' \right)
\end{equation}

The zero--mode truncated solution to~(\ref{bulkbrane}) is then
\begin{equation}
h_{\mu \nu}^{(0)} \left( x , l \right) = \int d^4 x' d \theta' G_0 \left( x, l ; x', l' \right) \, \left[ -2 \delta( l' - R \bt)\,\frac{\cos \bt }{M^4}\, \Sigma_{\mu\nu} \left( x' \right) \right] =  - \frac{4 \pi R \beta}{M^4 \, V_2} \cos \bt \: \left(\partial^2\right)^{-1} \, \Sigma_{\mu \nu} \left( x \right)
\end{equation}
By decomposing $\Sigma_{\mu \nu} \,$, one finds the result~(\ref{hsplit}) given in the main text.

\section{Perturbation of the volume of the internal space and of the length of the string}
\label{appH}

We compute the volume of the internal space (more precisely, of the half space with $m > 0$) up to first order in the perturbations. As in the rest of the paper, we use Gaussian normal coordinates at the brane.
\begin{equation}
V_2 + \Delta V = \int d l d \phi \sqrt{-g} = V_2 + \int d l d \phi \, \cos \, \theta \left( \Phi + C \right)
\end{equation}
where the background volume $V_2$ was also computed in~(\ref{v2}). Making use of eqs.~(\ref{invariant1}) and~(\ref{eins7}), we have
\begin{equation}
\Delta V = 2 \pi \beta R \left[ - 2 \int d l \, \cos \theta \, \Psi + \int d l \frac{d}{d l} \left( \cos \theta \, {\hat \zeta} \right) \right]
\end{equation}
The first integral can be computed separately on the outside and inside bulk. The second term is instead
a total derivative which can be rewritten as a boundary term at the string,
\begin{equation}
\int d l \frac{d}{d l} \left( \cos \theta \, {\hat \zeta} \right) = \cos \bt \left[ {\hat \zeta} \right]_J
\end{equation}
The sum of these terms gives
\begin{equation}
\Delta V =  2 \pi R^2\beta\,D_\Psi^{(i)} \, \frac{ 1 - \sin\bt }{ 3 }\,\left [6\, \beta + \left( 10 - 4\, \beta \right) \sin \bt - \left( 11 + \beta \right) \, \sin^2\bt \left(1+ \sin\bt\right)\right]
\end{equation}
This explicitly shows that the perturbation of the volume depends only on the $D_{\Psi}^{(i)}$ mode.

A similar conclusion is reached also for the perturbation of the circumference of the string,
\begin{eqnarray}
L &=& \int d \phi \sqrt{g_{\phi \phi}} = 2 \pi R \beta \cos \bt \left( 1 + C \right) = \nonumber\\
&=& 2\pi R\beta\,\cos\bt\,\left\{1+ D_\Psi^{(i)} \frac{1}{3\, \left( 1 - \beta \right)}\,\left[10\, \left( 1 - \beta \right) -9\, \left( 1 - \beta \right) \, \sin\bt + \left( 11 + \beta \right) \, \sin^3\bt\right]\right\}\,\,.
\end{eqnarray}

\bigskip

\bigskip

\footnotesize
\begin{multicols}{2}

\end{multicols}
\end{document}